\documentclass [12pt]{article}
\usepackage{graphicx,epsfig}

\newcommand{\be}{\begin{equation}}
\newcommand{\ee}{\end{equation}}
\def\ba{\begin{eqnarray}}
\def\ea{\end{eqnarray}}
\def\ni{\noindent}
\def\bib{\bibitem}
\def\del{\partial}
\def\le{\left}
\def\ri{\right}

\begin{document}
\baselineskip .75cm

\begin{center}
{\bf \Large Thermodynamics of Quasi-Particles}
\end{center}

\vspace{0.25 in}
\begin{center}
{F. G. Gardim$^{a}$ and  F. M. Steffens $^{a,b}$\\}
\vspace{0.25 in}
{\it \small \em $^{a}$ Instituto de F\'isica Te\'{o}rica -
Universidade
Estadual Paulista,\\
Rua Pamplona 145, 01405-900, S\~ao Paulo, SP, Brazil.\\}

\vspace{0.5cm}

 {\it \small \em $^{b}$ NFC - CCH - Universidade
Presbiteriana Mackenzie,\\
Rua da Consola\c{c}\~ao 930, 01302-907, S\~ao Paulo, SP, Brazil.}

\end{center}
\begin{abstract}

We present in this work a generalization of the solution of
Gorenstein and Yang for a consistent thermodynamics for systems with
a temperature dependent Hamiltonian. We show that there is a large
class of solutions, work out three particular ones, and discuss
their physical relevance. We apply the particular solutions for an
ideal gas of quasi-gluons, and compare the calculation to lattice
and perturbative QCD results.

\end{abstract}
\vspace{1cm}

\noindent

\newpage

\section{Introduction}

Lattice QCD suggests \cite{khan,karsch,cheng} that at sufficiently
high temperature $T$ and/or quark chemical potential $\mu$, the
strongly interacting matter exhibits a transition from a hadronic
phase to a new state, where matter is described in terms of
fundamental gluon and quark degrees of freedom \cite{collins}, the
Quark-Gluon Plasma (QGP). These lattice simulations suggest that
the transition occurs for temperatures around $T_c \sim 190$ MeV
\cite{cheng}. Heavy-ion collisions at RHIC, and in the future at the
LHC, provide us essential experimental data to the quest of the QGP
at high temperatures and small chemical potential. This experimental
search of the QGP needs reliable theoretical estimates of various
quantities, such as pressure, entropy, deconfinement temperature and
equations of state (EoS).

Perturbative QCD at finite $T$ and $\mu$ is one of the theoretical
tools to compute the various QGP quantities. However, strict
perturbation theory, which have been pushed up to $g_s^6\ln(1/g_s)$
\cite{kajantie}, is reasonable only for extremely high temperatures:
at temperatures near $T_c$ it is, in principle, not applicable, and
further treatments appear to be necessary. Moreover, the
perturbative series seems to be weakly convergent
\cite{kajantie,arnold}. Specifically, it is expected that when $T\gg
T_c$, the plasma behaves like an ideal gas of quarks and gluons, but
the perturbative series goes very slowly towards this expectation. A
way to attack this problem is through the reorganization of the
perturbative series. For instance, there are attempts using
resummation based on the Hard-Thermal-Loop (HTL) effective action
\cite{braaten}, in the form of so-called HTL perturbation theory
\cite{andersen1,andersen2}, and also attempts based on the 2-loop
$\Phi$-derivable approximation \cite{blaizot,blaizot3}. The latter
approach, which assumes a massive quasi-particle formalism
\cite{pe.1,levai}, leads to results which agree remarkably well with
lattice data.

The quasi-particle model of the Quark-Gluon plasma (qQGP) is a
phenomenological model which assumes non-interacting massive
quasi-particles which, with the aid of few parameters, like thermal
masses, is able to fit lattice QCD data over a wide range of
temperatures \cite{peshier,s.1,rebhanroma}: not only at extremely
high temperatures as in strict perturbation theory, or when
$T>20T_c$ as in the HTL perturbation theory, but also near $T_c$.
Quasi-particle models intend to describe deconfined matter from
$T_c$ up to $T\rightarrow\infty$. In the usual quasi-particle model
for the quark-gluon plasma \cite{pe.1,peshier}, statistical
mechanics for an ideal massive gas is used, where the mass of each
quasi-particle is dependent on the temperature (and on the chemical
potential) of the gas. However, when this idea is naively applied
there appear inconsistencies in the  thermodynamics of the system
\cite{go.1}. The in depth study of a general solution for the
thermodynamics of a system composed of particles with thermal masses
is the main objective of the present work.

In Sec. $2$ we will formulate the requirements for thermodynamics
self-consistency through the use of statistical mechanics in the
canonical ensemble, and explicitly compute three solutions. In Sec.
$3$ we will use one of these solutions to solve the problem for an
ideal gas of quasi-gluons, and obtain the generalized thermodynamics
relations as a function of the temperature. In Sec. $4$ we discuss
which solutions have physical meaning, and in Sec. $5$ we analyze
and summarize our results.

\section{Statistical Mechanics of Canonical Ensemble with a Temperature Dependent Hamiltonian}

The partition function of a  classical system in a canonical
ensemble whose Hamiltonian depends on an extra parameter T, which
will be identified with the temperature of the system, has the same
formal structure of the partition function of a system that has a T
independent Hamiltonian. Specifically:

\be Q_N\equiv \int \frac{dpdq}{N!(2\pi)^{3N}}e^{-\beta H},
\label{q-n} \ee

\ni where $\beta\equiv T^{-1}$, $N$ is the number of particles, $H$
is the system Hamiltonian which depends on the momentum $p$, the
coordinates $q$ of each particle, and the temperature of the system.
The integral is computed over all momenta and coordinates of the
particle (we use the notation $\int dpdp\equiv\int d^{3N}pd^{3N}q$).

The thermodynamics of the system is obtained from the partition
function

\be A(V,T)=-T\ln Q_N, \label{q} \ee

\ni where $A$ is a thermodynamic function to be determined later.
For instance, in the Standard Statistical Mechanics ($SM$), one
works with a temperature independent Hamiltonian and $A(V,T)$ is the
free energy.

In the $SM$, all the other thermodynamics functions can be found
from $A(V,T)$ using the thermodynamics relations:

\ba
P&=&-\le(\frac{\del A}{\del V}\ri)_T, \nonumber\\
S&=&-\le(\frac{\del A}{\del T}\ri)_V, \nonumber\\
U&=&A+TS, \label{A-relations} \ea

\ni where $P$ is the pressure, $S$ is the entropy, and $U$ is the
internal energy.

Returning to the case of a temperature dependent Hamiltonian, it
can be deduced from Eqs. (\ref{q-n}) and (\ref{q}) the following
identity

$$ \frac{1}{N!(2\pi)^{3N}}\int dpdqe^{\beta[A(V,T)-H(p,q,T)]}=1. $$

\ni Differentiating it with respect to $\beta$ on both sides, and
averaging the resulting expression, one obtains

\be A(V,T)-T\frac{\del A(V,T)}{\del T}-\langle
H(p,q,T)\rangle+T\Bigg\langle\frac{\del H(p,q,T)}{\del
T}\Bigg\rangle=0. \label{A?} \ee

\ni The symbol $\langle g\rangle$ means the average of the
function $g$ in the $SM$, and it is defined as

$$\langle g\rangle=(2\pi)^{3N}N!\int
dqdp g(q,p)\rho,$$

\ni where $\rho$ is the ensemble distribution function, given by

$$\rho=\frac{e^{-\beta H}}{Q_N}.$$

In the $SM$ the last term of the Eq. (\ref{A?}) vanishes, and the
internal energy is assumed to be $U=\langle H\rangle$. Then,
comparing equations (\ref{A-relations}) and (\ref{A?}), one
concludes that $A$ has to be the free energy, as stated before. As
the last term of Eq. (\ref{A?}) does not vanish in general, the
usual definition can not be used here. Thus, some questions appear
for a temperature dependent Hamiltonian: which thermodynamic
function is $A$? Has $U$ the same definition as in the $SM$?

We will show that there are innumerable possible answers to these
questions, with the constraint that one always has to recover the
standard Statistical Mechanics in the limit of a  $T$ independent
Hamiltonian. For any of the possible solutions, it is necessary to
redefine the connection between statistical mechanics and
thermodynamics, in such a way that the thermodynamics of the system
is also consistently built.

As Eq. (\ref{A?}) is the one where the problem is explicit, one
should start from it. First, one can redefine the free energy as a
function of the $SM$ free energy $A$:

\be A'(V,T,f(T))=A(V,T,f(T))+\alpha B(V,T), \label{A-redef} \ee

\ni where $B\equiv B(V,T)$ is an extra term which will be chosen to
let the thermodynamics formulation consistent, $\alpha$ is an
arbitrary real constant, and $f(T)$ is the additional temperature
contribution to the thermodynamics functions, which appears because
of the $T$ dependent Hamiltonian. In the limit of a T independent
Hamiltonian, $B(T)$ has to be zero, so one recovers all expressions
of the $SM$. The redefined  free energy $A'$ has to satisfy all the
thermodynamics relations, meaning that when replacing  Eq.
(\ref{A-redef}) into Eq (\ref{A-relations}), one finds:

\ba
P'&=&-\le(\frac{\del A'}{\del V}\ri)_T, \\
S'&=&-\le(\frac{\del A'}{\del T}\ri)_V, \\
\nonumber \label{A'-relations} \ea

\ni and

\be A'=U-TS+\alpha B. \label{eqa} \ee

\ni The quantities without prime are those calculated in the $SM$,
i.e. they are calculated regarding  $f(T)$  as a constant. In order
to let Eq. (\ref{eqa}) to have the same form of the third
thermodynamic relation given by Eq. (\ref{A-relations}), it is
necessary to redefine $U$ and $S$:

\ba
S'(V,T,f(T))=S(V,T,f(T))-\gamma \frac{B(V,T)}{T},\nonumber\\
U'(V,T,f(T))=U(V,T,f(T))+\eta B(V,T), \label{S-U-redef} \ea

\ni where $\gamma$ and $\eta$ are arbitrary constants, but
constrained by $\alpha=\gamma+\eta$. Using the second relation of
Eq. (\ref{A-relations}) and Eqs. (\ref{A-redef}) and
(\ref{S-U-redef}), one obtains

$$\gamma\frac{B}{T}=\alpha\frac{\del B}{\del
T}+\frac{df}{dT}\le(\frac{\del A}{\del f}\ri)_{V,T},
$$

\ni and manipulating the partial differential equation, one finds
two possible solutions,

\ba \frac{\del }{\del
T}(BT^{-\frac{\gamma}{\alpha}})&=&-\frac{T^{-\frac{\gamma}{\alpha}}}{\alpha}\frac{df}{dT}\le(\frac{\del
A}{\del
f}\ri)_{V,T} \hspace{1cm}\alpha\neq 0, \nonumber\\
B&=&\frac{T}{\gamma}\frac{df}{dT}\le(\frac{\del A}{\del
f}\ri)_{V,T}\hspace{2.3cm}\alpha=0. \label{B.e.d.A} \ea

\ni Using the definitions of $U'$, $S'$ and $A'$ in Eq.
(\ref{A?}), one obtains

$$\gamma B=T\alpha\frac{\del B}{\del
T}+T\Bigg\langle\frac{\del H}{\del T}\Bigg\rangle,
$$

\ni or,

\ba \frac{\del }{\del
T}(BT^{-\frac{\gamma}{\alpha}})&=&-\frac{T^{-\frac{\gamma}{\alpha}}}{\alpha}\Bigg\langle\frac{\del H}{\del T}\Bigg\rangle \hspace{1cm}\alpha\neq 0, \nonumber\\
B&=&\frac{T}{\gamma}\Bigg\langle\frac{\del H}{\del
T}\Bigg\rangle\hspace{2.3cm}\alpha=0. \label{B.e.d.H} \ea

As the entropy in the $SM$ can be written as

$$S=\beta\langle H\rangle+\ln Q_N=-\langle \ln\rho\rangle, $$

\ni so, from Eq. (\ref{S-U-redef}), one is able to find the
connection between the ensemble distribution function and the
redefined entropy:

\be S'(V,T,f(T))=-\langle \ln\rho\rangle-\gamma \frac{B(V,T)}{T}.
\label{S-rho} \ee

\ni With the redefined thermodynamics functions, one then constructs
a general model to treat the case of a $T$ dependent Hamiltonian,
where all the thermodynamics relations are satisfied. Hamiltonian
functions which depends on the temperature of the system appears in
problems with mean field approximation, as in the theory of nuclear
matter \cite{walecka}, or in phenomenological models of Quark Gluon
Plasma \cite{pe.1,peshier,go.1}. One is able to recover the
thermodynamics consistency of systems with a  $T$-dependent
Hamiltonian if proper care of the extra term $B$ is taken. The
natural question is: what is the meaning of $B$? As the $N!$ factor
in Eq. (\ref{q-n}) can not be justified classically, the meaning of
$B$  is to be interpreted with the aid of quantum mechanics.

For a quantum system, the canonical ensemble with a Hamiltonian
operator $\hat{H}$ has the following definition:

\ba
Q_N &\equiv & Tr e^{-\beta \hat{H}},\nonumber\\
\hat{\rho}&=&\frac{e^{-\beta \hat{H}}}{Q_N},\nonumber\\
\langle \hat{g}\rangle&=&Tr(\hat{g}\hat{\rho}), \label{q-n-Q} \ea

\ni where $\hat g$ is a given hermitian operator and $\hat{\rho}$ is
the ensemble density operator. Based in the redefined classical
internal energy Eq. (\ref{S-U-redef}), one is able to write the
redefined quantum internal energy, with the help of Eq.
(\ref{q-n-Q}), as

$$U'=\langle \hat{H}\rangle+\eta B=Tr[\hat{\rho}(\hat{H}+\eta\hat{B})],$$

\ni with $\hat{B}=B\textbf{1}$, where $\textbf{1}$ is the unitary
matrix. Comparing this equation for $U'$ with Eq. (\ref{q-n-Q}), one
is lead to assume the term inside the brackets as the total
Hamiltonian

\be \hat{H}_T=\hat{H}+\hat{E}_0, \label{H-Total} \ee

\ni where $\hat{E}_0=\eta\hat{B}$. As $B$ does not depend on the
momenta and coordinates, the density operator for a general
Hamiltonian $\hat{H}_T$ has exactly the same form as that given by
Eq. (\ref{q-n-Q}). Also, as $\hat{\rho}$ does not change for the
redefined Hamiltonian, the entropy does not change as well. The
quantum statistical mechanics relations are then written as

\ba S'&=&-\langle \ln\hat{\rho}\rangle-\gamma
\frac{B}{T}\nonumber\\
A'&=&-T\ln Q^T_N+\gamma B\nonumber\\
U'&=&\langle\hat{H}_T\rangle. \label{finalexp-SUA} \ea

\ni Here, $Q^T_N$ is the partition function written in terms of
$\hat{H}_T$. Note that for $\gamma=0$, the thermodynamics relations
have the same form as those of the $SM$. The interpretation for $B$
was first given by Gorenstein and Yang in Ref. \cite{go.1}, where
they make the observation that in the standard case of a $T$
independent Hamiltonian, the zero point energy is a constant and it
is usually subtracted out because experiments measure only energy
differences. In the quasi-particle model, the dispersion relation is
$T$ dependent, and so is the zero point energy of the system. Thus,
it can not be discarded from the energy spectrum. In this sense,
$\eta B$ is the system energy in the absence of quasi-particle
excitations, i.e. the system lowest state energy. Note from Eqs.
(\ref{B.e.d.A}), (\ref{B.e.d.H}) and (\ref{finalexp-SUA}), that when
one of the three constants, $\alpha$, $\gamma$ or $\eta$ are zero,
the corresponding thermodynamics relations are independent of the
two remaining constants, i.e., they cancel each other.

In this section, it was determined all possible mathematical
solutions for the formulation of a consistent thermodynamics for
systems with a $T$ dependent Hamiltonian. The constants $\eta$ and
$\gamma$ can assume any value, consequently $\alpha$ as well, but
some of the values of these constants are of practical use in
physics. In the next subsections, three of these particular
situations will be developed.


\subsection{Solution 1}

The first solution to be dealt with is the one with $\gamma=0$ and
$\alpha=\eta$, which implies that entropy is not changed, while the
free energy and the internal energy are:

\ba
A'&=&-T\ln Tr\le(e^{-\beta \hat{H}_T}\ri)=-T\ln Tr\le(e^{-\beta \hat{H}}\ri)+B\nonumber\\
U'&=&\frac{1}{Q_N^T}Tr\le(\hat{H}_Te^{-\beta
\hat{H}_T}\ri)=\frac{1}{Q_N}Tr\le(\hat{H}e^{-\beta \hat{H}}\ri)+B.
\label{AU-1} \ea

\ni From Eqs. (\ref{B.e.d.A}) and (\ref{B.e.d.H}), one determines
the $B$ term:

\be B=B_0-\int dT\Bigg\langle\frac{\del \hat{H}}{\del
T}\Bigg\rangle, \label{B1} \ee

\ni where $B_0$ is an integration constant. Using this equation for
$B$, the redefined thermodynamics functions are

\ba
S'&=&-\langle \ln \hat{\rho}\rangle, \nonumber\\
A'&=&-T\ln Q_N-\int dT\Bigg\langle\frac{\del \hat{H}}{\del
T}\Bigg\rangle+B_0, \nonumber\\
U'&=&\langle \hat{H}\rangle-\int dT\Bigg\langle\frac{\del
\hat{H}}{\del T}\Bigg\rangle+B_0. \label{AUS-1-final} \ea

This solution was first used by Gorenstein and Yang \cite{go.1} to
study a gluon plasma with a $T$ dependent Hamiltonian, where the
gluon dispersion relation was assumed to have a $T$ dependent gluon
mass. In that case, the $f(T)$ function is given by the square of
the quasi-gluon mass. This solution was used in several works, where
the quark-gluon plasma is treated by a quasi-particle model
\cite{pe.1,peshier,s.1,rebhanroma}, and it can be physically
supported: if one views a hadron through a bag model, then one deals
with a ``bag pressure'' and a ``bag energy". When heat is added to
the system, there will be some reminiscence from the bag on the
energy and on the pressure: $B$ plays this role - generally $B$ is
referred to as  bag energy or bag pressure.

\subsection{Solution 2}

One other possible solution is $\eta=0$ and $\alpha=\gamma$. Then,
according to Eq. (\ref{finalexp-SUA}), the internal energy is
unchanged, and the entropy and the free energy are given by:

\ba
A'&=&-T\ln Tr\le(e^{-\beta \hat{H}_T}\ri)+B=-T\ln Tr\le(e^{-\beta \hat{H}}\ri)+B\nonumber\\
S'&=&\frac{1}{Q_N^T}Tr\le(e^{-\beta
\hat{H}_T}\ln\hat{\rho}_T\ri)-\beta B=\frac{1}{Q_N}Tr\le(e^{-\beta
\hat{H}}\ln\hat{\rho}\ri)-\beta B. \label{AS-2} \ea

Eqs. (\ref{B.e.d.A}) and (\ref{B.e.d.H}) can be rewritten for
$\eta=0$ as:

\be \frac{\del}{\del
T}\le(\frac{B}{T}\ri)=-\frac{1}{T}\frac{df}{dT}\le(\frac{\del
A}{\del f}\ri)_{V,T}=-\frac{1}{T}\Bigg\langle\frac{\del\hat{H}}{\del
T}\Bigg\rangle \label{B2-1}. \ee

\ni Computing $B$ from Eq. (\ref{B2-1}), the redefined
thermodynamics functions are

\ba S'&=&-\langle \ln \hat{\rho}\rangle+\int
dT\frac{1}{T}\Bigg\langle\frac{\del \hat{H}}{\del T}\Bigg\rangle-\frac{B_0}{T_0},\nonumber\\
A'&=&-T\le(\ln Q_N+\int dT\frac{1}{T}\Bigg\langle\frac{\del\hat{H}}{\del T}\Bigg\rangle-\frac{B_0}{T_0}\ri), \nonumber\\
U'&=&\langle \hat{H}\rangle. \label{ASU-2-final} \ea

This second solution was used in Ref. \cite{Bannur}. It describes
the quark-gluon plasma by a quasi-particle model satisfying all
thermodynamics relations as well. However, here the physics
motivation is that the whole interaction energy goes to the
quasi-particle mass, i.e. there is no extra term in the usual
internal energy expression.

\subsection{Solution 3}

A third solution consists in $\alpha=0$ and $\eta=-\gamma$. This is
simpler than the others, since with $\alpha=0$, Eqs. (\ref{B.e.d.A})
and (\ref{B.e.d.H}) for $B$ are not partial differential equations.
In this solution, the free energy is unchanged, while the entropy
and the internal energy are:

\ba
S'&=&\frac{1}{Q_N^T}Tr\le(e^{-\beta\hat{H}_T}\ln\hat{\rho}_T\ri)+\beta B=\frac{1}{Q_N}Tr\le(e^{-\beta\hat{H}}\ln\hat{\rho}\ri)+\beta B, \nonumber \\
U'&=&\frac{1}{Q_N^T}Tr\le(\hat{H}_Te^{-\beta
\hat{H}_T}\ri)=\frac{1}{Q_N}Tr\le(\hat{H}e^{-\beta \hat{H}}\ri)+B.
\label{AS-3} \ea

The redefined thermodynamics functions are then

\ba
S'&=&-\langle\ln\hat{\rho}\rangle-\Bigg\langle\frac{\del\hat{H}}{\del
T}\Bigg\rangle, \nonumber\\
A'&=&-T\ln Q_N,\nonumber\\
U'&=&\Bigg\langle\frac{\del(\beta \hat{H})}{\del
\beta}\Bigg\rangle \label{AUS-3-final}. \ea

\ni Note that the internal energy in Eq. (\ref{AUS-3-final}) is the
only one of the three solutions in which the connection between the
partition function and $U'$ has the same form as in the $SM$ case
\cite{huang}:

$$U'=-\frac{\del}{\del\beta}\ln Q_N.$$

In the quasi-particle model context of the quark-gluon plasma, the
third solution can be interpreted as the following: the
quasi-particle changes the zero point energy, such that it becomes a
function of $T$ (just as the quasi-particle mass). This effect
occurs because the whole interaction energy can not be accommodated
in the quasi-particle mass, and part of it is used by the vacuum to
modify its zero point energy, represented here by $B(T)$. In this
solution, the entropy definition is also modified, which may be seen
as a problem. However, as it is well known, the entropy of a system
is determined up to a constant that is usually subtracted from the
system entropy. This procedure can not be done here, because for the
case of a $T$ dependent dispersion relation, the additive term to
the entropy will not be a constant, but also a function of the
temperature.


\section{Ideal Quasi-Particle Gas Model}

In the previous section, we have constructed a self-consistent
thermodynamics for a system with a  $T$ dependent Hamiltonian. In
this section we will apply this theory for one specific case: a
plasma composed only by gluons, with a vanishing chemical potential.
The quasi-particle model implies that we will treat the plasma as a
non-interacting gas of massive, temperature dependent, gluons. This
problem has already been solved for two particulars solutions,
solution $1$ \cite{go.1} and $2$ \cite{Bannur} of the previous
section, of the general solution given there. To exemplify our
method we choose the easiest solution, solution 3, which gives the
simplest extra term $B$. We will see that we can obtain an algebraic
solution for the thermodynamics function, differently from solutions
1 and 2.

\subsection{The $\alpha=0$ Solution}

It is interesting to work in a \textit{grand canonical ensemble}
because the thermodynamics functions are easier to compute there,
but we have developed the theory in a canonical ensemble.
Nevertheless, for the $\mu=0$ case it is not necessary to redefine
the whole theory in the grand canonical ensemble, because all that
is required for this change of ensemble is to use the grand
canonical partition function, $Z$, instead of $Q_N$. This connection
is justified because both ensembles are equivalent in the
thermodynamical limit, and for $\mu=0$ the additional thermodynamics
relation for $N$, $N=-\del A/\del \mu$, disappears. The grand
partition function is given by:

\be Z(z,V,T)\equiv\sum_{N=0}^\infty z^NQ_N(V,T), \label{Z} \ee

\ni where we introduced the fugacity $z$,

\be z=e^{\beta\mu}. \label{fugacity} \ee

\ni The distribution function $\rho$ is

$$\hat{\rho}=\frac{e^{-\beta\hat{H}}}{Z},$$

\ni and the mean value is defined as

$$\langle A\rangle=\frac{\sum_{N=0}^\infty Az^NQ_N(V,T)}{\sum_{N=0}^\infty z^NQ_N(V,T)}.$$

\ni Using the results given in the Sec. 2.3, one has the redefined
thermodynamics functions:

\ba
S'&=&-\langle\ln\hat{\rho}\rangle-\Bigg\langle\frac{\del\hat{H}}{\del
T}\Bigg\rangle, \nonumber\\
P'V&=&T\ln Z,\nonumber\\
U'&=&\Bigg\langle\frac{\del(\beta \hat{H})}{\del
\beta}\Bigg\rangle=-\frac{\del}{\del\beta}\ln Z,
\label{Z-sua-solution} \ea

\ni where the thermodynamic relation $A=-PV$ was used. From quantum
statistical mechanics, the partition function for a free gas is

$$ Z(V,T)=\prod_k\frac{1}{1-e^{-\beta\omega_k}},$$

\ni or

$$ \ln Z=-\sum_k\ln (1-e^{-\beta\omega_k})=-\frac{V}{(2\pi)^3}\int4\pi
dkk^2\ln (1-e^{-\beta\omega}). $$

\ni where the dispersion relation is $\omega^2=k^2+m^2(T)$.
Therefore $P'$ and $U'$ are given by Eqs. (\ref{Z-sua-solution}),
and take the form

\ba P'(T)=-\frac{\nu T}{2\pi^2}\int dkk^2\ln (1-e^{-\beta\omega})=
\frac{\nu}{6\pi^2}\int_0^{\infty}dkf(k)\frac{k^4}{\sqrt{k^2 +
m^2(T)}} \label{p}\\
e'(T)=\frac{U'}{V}=\frac{\nu}{2\pi^2}\int_0^{\infty}dkf(k)k^2\sqrt{k^2
+ m^2(T)}-\frac{\nu T}{4\pi^2}\frac{\partial m^2}{\partial
T}\int_0^{\infty}dkf(k)\frac{k^2}{\sqrt{k^2 + m^2(T)}}, \label{e}
\ea

\ni where $\nu$ is the gluon degeneracy factor, $e'(T)=\frac{U'}{V}$
is the energy density and $f(k)=(e^{\beta \omega}-1)^{-1}$ is the
Bose-Einstein distribution function. Note the second term in the
expression for the energy. It exists only because $m = m(T)$, and it
will be denoted by $b(T)\equiv B(T,V)/V$.

Using the relation $S'T=U'+P'V$, one obtains:

\be
s'(T)=\frac{\nu}{6\pi^2T}\int^{\infty}_{0}dkf(k)k^2\frac{4k^2+3m^2(T)}{\sqrt{k^2
+ m^2(T)}} -\frac{\nu}{4\pi^2}\frac{\partial m^2}{\partial
T}\int^{\infty}_{0}dkf(k)\frac{k^2}{\sqrt{k^2 + m^2(T)}}.
\label{s}
\ee

If one uses the thermodynamic relation  $S'=-[\del(P'V)/\del T]_V$
to calculate the entropy density, with the help of $P'$ given by
Eq. (\ref{p}), one finds exactly the same expression Eq. (\ref{s})
for $s'$, confirming that the calculation is self-consistent; the
additional term $b(T)$ in $e'(T)$ and $s'(T)$ guarantees the
thermodynamical consistency of the qQGP expressions. If one
compares this extra term with the usual extra term of qQGP, the
Gorenstein-Yang solution (Eq.(5) of Ref. \cite{peshier}), one
finds the relation $b(T)=\del B(T)/\del T$. This implies that it
is necessary to compute an integral in $T$ in order to get the bag
constant $B(T)$. In the present approach, there is not an integral
in $T$, what facilitates analytical calculations. Also, as $m(T)$
depends on the coupling constant, in the form of a logarithm in
$T$, $B(T)$ is harder to compute.


\subsubsection{Pressure}

The pressure $P'(T)$ is given by Eq. (\ref{p}), and it can be
rewritten as

\be
P'(T)=\frac{\nu\beta^{-4}}{6\pi^2}\int^{\infty}_{0}
dx\frac{x^4}{\sqrt{x^2+r^2}}\frac{1}{e^{\sqrt{x^2+r^2}}-1}=\frac{4\nu
T^4}{\pi^2}I_5(r) \label{pppp},
\ee

\ni where $r\equiv\frac{m}{T}$ and $I_5$ is Eq. (\ref{I5}) of {\it
Appendix A}. Eq. (\ref{pppp}) becomes

$$
P'(T)=\frac{\nu}{\pi^2}\left[\frac{\pi^4}{90}T^4-\frac{\pi^2}{24}T^2m^2+\frac{\pi}{12}Tm^3+\frac{m^4}{2^5}\left(\log\frac{m}{4\pi
T}+\gamma_E-\frac{3}{4}\right)+\frac{1}{8}\sum^{\infty}_{n=1}a_n\frac{m^{2(n+2)}}{T^{2n}}\right],$$

\ni with
$a_n=\frac{(-1)^n(2n-1)!!\zeta(2n+1)}{(n+2)!2^{3n+1}\pi^{2n}}$.
Rewriting the above integral as a function of the ideal massless gas
pressure, $P_{0}=\frac{\nu\pi^2T^4}{90}$, one obtains the pressure
for the massive gas of quasi-gluons:

\ba
P'(T)=P_{0}\left[1-\frac{15}{4\pi^2}\left(\frac{m}{T}\right)^2+\frac{15}{2\pi^3}\left(\frac{m}{T}\right)^3+\frac{45}{16\pi^4}\left(\frac{m}{T}\right)^4\left(\log\frac{m}{4\pi
T}+\gamma_E-\frac{3}{4}\right)+\right.\nonumber\\
\left.+\frac{45}{4\pi^4}\sum^{\infty}_{n=1}a_n\left(\frac{m}{T}\right)^{2(n+2)}\right].
\label{p(T)}
\ea

Note that the functional form of $m(T)$ is unknown, i.e. statistical
mechanics does not provide the function $m(T)$. This problem will be
treated on the next section.


\subsubsection{Energy}

The energy density $e'(T)$ is given by Eq. (\ref{e}), and using the
same technique as used in the calculation of the pressure, one is
able to get the expression for the energy density dependent on the
temperature:

\ba
e'(T)&=&\frac{\nu T^{4}}{2\pi^2}\int^{\infty}_{0}
dx\frac{x^4}{\sqrt{x^2+r^2}}\frac{1}{e^{\sqrt{x^2+r^2}}-1}+\frac{\nu
T^{2}}{2\pi^2}\le(m^2-\frac{T}{2}\frac{\del m^2}{\del
T}\ri)\int^{\infty}_{0}dx\frac{x^2}{\sqrt{x^2+r^2}}\frac{1}{e^{\sqrt{x^2+r^2}}-1}\nonumber\\
&=&\frac{12\nu T^4}{\pi^2}I_5(r)+\frac{\nu
T^{2}}{\pi^2}\le(m^2-\frac{T}{2}\frac{\del m^2}{\del T}\ri)I_3(r),
\label{eeee}
\ea

\ni where $I_3$ and $I_5$ are Eqs. (\ref{I3}) and (\ref{I5}) of {\it
Appendix A}. Solving these integrals one obtains:

\ba
e'(T)=e_0\left\{1-\frac{5}{4\pi^2}\le(\frac{m}{T}\ri)^2-\frac{15}{16\pi^4}\left(\frac{m}{T}\right)^4\left(\log\frac{m}{4\pi
T}+\gamma_E+\frac{1}{4}\right)-\frac{15}{4\pi^4}\sum^{\infty}_{n=1}a_n(2n+1)\left(\frac{m}{T}\right)^{2(n+2)}+\right.\nonumber\\
\left.-\frac{5}{4T\pi^2}\frac{\del m^2}{\del
T}\le[1-\frac{3}{\pi}\frac{m}{T}-\frac{3}{2\pi^2}\frac{m^2}{T^2}\left(\log\frac{m}{4\pi
T}+\gamma_E-\frac{1}{2}\right)-\frac{3}{\pi^2}\sum^{\infty}_{n=1}a_n(n+2)\left(\frac{m}{T}\right)^{2(n+1)}\ri]\right\}
.\label{e(T)} \ea

\ni where $e_0=\frac{\nu T^4 \pi^2}{30}$ is the energy density of
the ideal gas.


\subsubsection{Entropy}

As the expression for the pressure for the massive gas of
quasi-gluons has been obtained, it is easier to compute the entropy
density through the use of a Maxwell relation, and one gets:

\ba
s'(T)=\frac{\partial P'}{\partial
T}=\frac{\nu\pi^2}{90}\frac{\partial }{\partial
T}\left[T^4\left(1-\frac{15}{4\pi^2}\left(\frac{m}{T}\right)^2+\frac{15}{2\pi^3}\left(\frac{m}{T}\right)^3+\frac{45}{16\pi^4}\left(\frac{m}{T}\right)^4\left(\log\frac{m}{4\pi
T}+\gamma_E-\frac{3}{4}\right)+\right.\right.\nonumber\\
\left.\left.+\frac{45}{4\pi^4}\sum^{\infty}_{n=1}a_n\left(\frac{m}{T}\right)^{2(n+2)}\right)\right].\nonumber
\ea

\ni Using
$$\frac{\partial}{\partial T}\left(\frac{m}{T}\right)=
\frac{\partial m}{\partial T}\frac{1}{T}-\frac{m}{T^2}=\frac{1}{2mT}\frac{\partial m^2}{\partial T}-\frac{m}{T^2},$$

\ni $s'(T)$ is written as:

\ba
s'(T)=s_{0}\left\{1-\frac{15}{8\pi^2}\left(\frac{m}{T}\right)^2+\frac{15}{8\pi^3}\left(\frac{m}{T}\right)^3-\frac{45}{64\pi^4}\left(\frac{m}{T}\right)^4-\frac{45}{8\pi^4}\sum^{\infty}_{n=1}a_nn\left(\frac{m}{T}\right)^{2(n+2)}\right.\nonumber\\
\left.-\frac{15}{16T\pi^2}\frac{\del m^2}{\del
T}\le[1-\frac{3}{\pi}\frac{m}{T}-\frac{3}{2\pi^2}\frac{m^2}{T^2}\left(\log\frac{m}{4\pi
T}+\gamma_E-\frac{1}{2}\right)-\frac{3}{\pi^2}\sum^{\infty}_{n=1}a_n(n+2)\left(\frac{m}{T}\right)^{2(n+1)}\ri]\right\}.
\label{s(T)} \ea

\ni where $s_0=\frac{2\nu\pi^2T^3}{45}$ the ideal gas' entropy
density.


\subsubsection{Results}

The temperature dependence of the thermodynamics functions of a
quasi-particle gas can not be completely described using statistical
mechanics only. In particular, to determine the expression for
$m(T)$ it is necessary some further information. First, we take HTL
perturbation theory \cite{blaizot,rebhanroma} as the additional
information. HTL provides a relation between $m(T)$ and $T$ that,
asymptotically, is

\be m^2(T)=N_c\frac{g_s^2T^2}{6}, \label{m-assim} \ee

\ni where $g_s$ is the strong coupling constant. With this input,
one can compute the explicit temperature dependence of the
thermodynamics functions. As we are interested not just in the
leading order behavior but also in the higher order corrections, we
will implement these higher order corrections to the thermal mass of
the HTL approach \cite{blaizot},

\be \delta m^2=-N_c\frac{g_s^2}{2\pi}Tm_D, \label{m-m_D} \ee

\ni where $m_D=\sqrt{\frac{2N_c}{6}}g_sT$ is the Debye mass. For the
values of the strong coupling constant at high temperatures, $g_s
\ll 1$, expression (\ref{m-m_D}) should describe correctly the
next-to-leading order temperature dependence of the mass. However
this equation runs into problems, giving a tachyon mass for the
gluons, depending on the value of $g_s$. As the aim is to construct
a self-consistent model to describe the deconfined QCD phase using
quasi-particles, we can not take this next-to-leading mass as the
complete quasi-gluon asymptotic mass but as an approximation to the
complete mass. Analyzing this expression inside the HTL picture
\cite{blaizot}, it is seen that Eq. (\ref{m-m_D}) comes exclusively
from hard momenta corrections, i.e. soft momenta corrections do not
enter. The complete description of the asymptotic mass term requires
both soft and hard corrections, and these will be simulated through
a quadratic gap equation given by,

\be m^2(T)=N_c\frac{g_s^2T^2}{6}-\frac{N_c}{\sqrt{2}\pi}g_s^2Tm(T).
\label{m-NLO} \ee

\ni Using Eq. (\ref{m-NLO}) in Eq. (\ref{p(T)}), one is able to
compare the present results with lattice QCD data \cite{karsch}. For
the calculation of the temperature dependence of the coupling
constant, we used \cite{pks}:

\be g_s^2(T)=\frac{48\pi^2}{11N_c\log[\lambda T/T_c+T_s/T_c]^2},
\label{g(T)} \ee

\begin{figure}[h]
\begin{center}
\includegraphics[height=7cm,width=12cm]{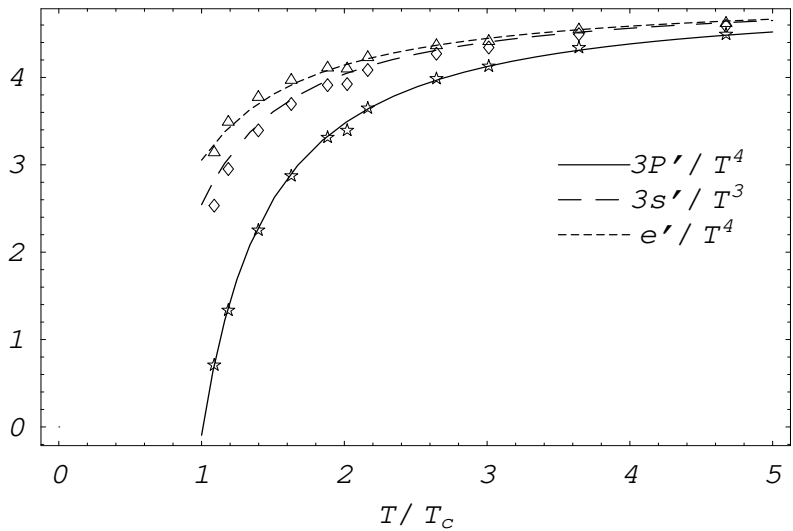}
\end{center}
\caption {Plots of the pressure, energy density and entropy
density in our model, and the extrapolated lattice data
\cite{karsch} for the pressure (star), energy density (triangle)
and entropy density (diamond), as a function of $T/T_c$.}
\label{fig1}
\end{figure}

\ni with $\lambda$ and $T_s$  phenomenological parameters, and
$N_c=3$ is the number of colors that will be used hereafter. The
optimal fit is achieved using $\lambda = 5.182$, $T_s/T_c = -0.197$,
and the degeneracy factor $\nu=2(N_c^2 - 1)=16$. In Fig. \ref{fig1}
is plotted the thermodynamics functions Eqs. (\ref{p(T)}),
(\ref{e(T)}) and (\ref{s(T)}), using the mass relation Eq.
(\ref{m-NLO}). The calculated curves describe quite well the lattice
data \cite{karsch}. For comparison, in Ref. \cite{peshier}, where
the qQGP was based on the Gorenstein and Yang solution, it was
necessary 4 parameters to get an optimal fit of the lattice data:
$\lambda$, $T_s/T_c$, the integration constant $B_0$ and the
degeneracy $\nu$.

\subsection{Asymptotic Behavior of the $\alpha \neq 0$ Solutions}

In the last section, the $\alpha=0$ solution was studied in some
detail using explicit expressions. For the $\alpha\neq 0$ cases,
however, the calculation of the $B$ term involves non trivial
integrals. Nevertheless, if one is interested only in the
asymptotic behavior of the gluon plasma, an explicit calculation
can also be done. In this sub-section, we will focus on this
calculation.

The general expression for the $\alpha\neq 0$ solution for $B$ is
given by Eq. (\ref{B.e.d.H}). Using the quasi-gluon dispersion
relation, one can rewrite $B$ as

\be B=B_0T^{\frac{\gamma}{\alpha}}-\frac{\nu
VT^{\frac{\gamma}{\alpha}}}{\alpha
4\pi^2}\int^T_{T_c}d\tau\tau^{-\frac{\gamma}{\alpha}}\frac{dm^2(\tau)}{d\tau}\int^{\infty}_0dk
\frac{k^2f(k)}{\sqrt{k^2+m^2(\tau)}}. \nonumber \label{eqB} \ee

\ni To analyze the high $T$ behavior of $B(T)$, we manipulate Eq.
(\ref{eqB}) in order to use the form found in Appendix A:

\ba
B&=&B_0T^{\frac{\gamma}{\alpha}}-\frac{\nu
VT^{\frac{\gamma}{\alpha}}}{\alpha
4\pi^2}\int^T_{T_c}d\tau\tau^{-\frac{\gamma}{\alpha}}\frac{dm^2(\tau)}{d\tau}\Gamma(3)\tau^2I_3\le(\frac{m}{\tau}\ri)\nonumber\\
&=&B_0T^{\frac{\gamma}{\alpha}}-\frac{\nu
VT^{\frac{\gamma}{\alpha}}}{\alpha
2\pi^2}\int^T_{T_c}d\tau\tau^{2-\frac{\gamma}{\alpha}}\frac{dm^2(\tau)}{d\tau}\le(\frac{\pi^2}{12}-
\frac{\pi}{4}\frac{m}{\tau}+\emph{O}\le(\frac{m^2}{\tau^2}\ri)\ri).
\label{Bin}
\ea

\ni Integrating and using just the first two terms inside the
parenthesis, one gets

\ba
B=B_0T^{\frac{\gamma}{\alpha}}-\frac{\nu
V}{24\alpha}\le[m^2\tau^2\Big|^T_{T_c}-\le(2-\frac{\gamma}{\alpha}\ri)T^{\frac{\gamma}{\alpha}}\int^T_{T_c}d\tau\tau^{1-\frac{\gamma}{\alpha}}m^2\ri]+\nonumber\\
+\frac{\nu
V}{12\alpha\pi}\le[m^3\tau\Big|^T_{T_c}-\le(1-\frac{\gamma}{\alpha}\ri)T^{\frac{\gamma}{\alpha}}\int^T_{T_c}d\tau\tau^{-\frac{\gamma}{\alpha}}m^3\ri]+...
\label{Bform} \ea

\ni On the other hand, for $g_s\ll 1$, Eq. (\ref{m-NLO}) is reduced
to

\be \frac{m}{T}=\frac{g_s}{\sqrt{2}}-\frac{3g_s^2}{2\sqrt{2}\pi}.
\label{m/T} \ee

\ni At very high temperatures, the coupling constant, Eq.
(\ref{g(T)}), decreases very slowly, implying that the ratio $m/T$
is a slowly decreasing function at $T/T_c\gg 1$. Hence, $m/T$ can be
regarded as a constant, and the integral can be easily done.
Rewriting the integrand in $m/T$ powers, one can compute the
integral of the $B$ term at very high temperatures:

\ba
B&\approx& B_0T^{\frac{\gamma}{\alpha}}-\frac{\nu
V}{24\alpha}\le[m^2T^2-\le(2-\frac{\gamma}{\alpha}\ri)T^{\frac{\gamma}{\alpha}}\frac{m^2}{T^2}\frac{T^{4-\frac{\gamma}{\alpha}}}{4-\frac{\gamma}{\alpha}}\ri]+\frac{\nu
V}{12\alpha\pi}\le[m^3T-\le(1-\frac{\gamma}{\alpha}\ri)T^{\frac{\gamma}{\alpha}}\frac{m^3}{T^3}\frac{T^{4-\frac{\gamma}{\alpha}}}{4-\frac{\gamma}{\alpha}}\ri]\nonumber\\
&=& B_0T^{\frac{\gamma}{\alpha}}-\frac{\nu
V}{12\alpha}\frac{1}{4-\gamma/\alpha}m^2T^2+\frac{\nu
V}{4\alpha\pi}\frac{1}{4-\gamma/\alpha}m^3T, \hspace{1cm}
\frac{\gamma}{\alpha}\neq 4. \label{Bdif4}
\ea

\ni Note that Eq. (\ref{Bdif4}) is not valid for
$\frac{\gamma}{\alpha}=4$. For $\frac{\gamma}{\alpha}=4$, one has

\ba
B\approx B_0T^{4}-\frac{\nu V}{24\alpha}(1+2\log
T)m^2T^2+\frac{\nu V}{12\alpha\pi}(1+3\log T)m^3T, \hspace{1cm}
\frac{\gamma}{\alpha}=4.
\label{Bequal4}
\ea

\ni As the asymptotic mass is proportional to $T$ in first order,
one has to maintain the $B_0$ term. To obtain the asymptotic
thermodynamics functions, we write the asymptotic $B$ in terms of
ideal massless gas pressure $P_0$:

\ba b&\approx&
P_0\le(\frac{90b_0}{\nu\pi^2}-\frac{15}{4\alpha\pi^2}(1+2\log
T)\frac{m^2}{T^2}+\frac{ 15}{2\alpha\pi^3}(1+3\log
T)\frac{m^3}{T^3}\ri), \hspace{.4cm} \frac{\gamma}{\alpha}=4; \nonumber\\
b &\approx&
P_0\le(b_0\frac{90}{\nu\pi^2}T^{\frac{\gamma}{\alpha}-4}-\frac{15}{2\alpha\pi^2}\frac{1}{4-\gamma/\alpha}\frac{m^2}{T^2}+\frac{
45}{2\alpha\pi^3}\frac{1}{4-\gamma/\alpha}\frac{m^3}{T^3}\ri),
\hspace{1cm} \frac{\gamma}{\alpha}\neq 4,
\label{Bapprox}
\ea

\ni where $b_0\equiv \frac{B_0}{V}$. The thermodynamics functions
can now be computed. For instance,  with the help of Eqs.
(\ref{A-redef}), (\ref{p(T)}) and (\ref{Bapprox}) one is able to
obtain the asymptotic expression for the pressure for $\alpha\neq 0$
and $\gamma/\alpha=4$:

\ba
P'(T)=P_{0}\left[1-\frac{90b_0\alpha}{\nu\pi^2}+\frac{15}{2\pi^2}\frac{m^2}{T^2}\log
T-\frac{45}{2\pi^3}\frac{m^3}{T^3}\log T+\ldots\ri].\nonumber \ea

\ni From Eq. (\ref{g(T)}), it is seen that the $T$ logarithm is
inversely proportional to $g_s^2$. Thus, with the help of the
asymptotic mass Eq. (\ref{m-assim}), one obtains the asymptotic
pressure:

\ba P'(T)=P_{0}\left[1-\frac{90b_0\alpha}{\nu\pi^2}+\frac{30}{11}+
\frac{90}{11\sqrt{2}}\frac{g_s}{\pi}+\ldots\ri].\nonumber \ea

\ni When the coupling constant goes to zero, $P'(T)$ should be that
of an ideal massless gas. To this aim, one has to choose
$b_0\alpha=\nu\pi^2/33$. The pressure is then

\ba P'(T)=P_{0}\left[1+
\frac{90}{11\sqrt{2}}\frac{g_s}{\pi}+\ldots\ri],\hspace{1cm}
\frac{\gamma}{\alpha}=4. \label{Passim4} \ea

\ni The pressure for $\gamma/\alpha\neq 4$ is easier to work with.
Using Eqs. (\ref{A-redef}), (\ref{p(T)}) and (\ref{Bapprox}) one
gets

\ba
P'(T)=P_{0}\left[1-\frac{15}{4\pi^2}\frac{2-\frac{\gamma}{\alpha}}{4-\frac{\gamma}{\alpha}}\le(\frac{m}{T}\ri)^2+\frac{15}{2\pi^3}\frac{1-\frac{\gamma}{\alpha}}{4-\frac{\gamma}{\alpha}}\le(\frac{m}{T}\ri)^3-
\alpha
b_0\frac{90}{\nu\pi^2}T^{\frac{\gamma}{\alpha}-4}+\ldots\right],\hspace{1cm}
\frac{\gamma}{\alpha}\neq 4. \label{Passim-dif4} \ea

In section II, the general mathematical solution for a consistent
thermodynamics of massive gluons was handled. We now want to address
the problem from the physics point of view: which solutions are
physically relevant?

The first feature that one has to keep in mind is that the gluon
plasma must behave as an ideal massless gas in the limit
$T\rightarrow\infty$. This property implies that the pressure has to
be proportional to $T^4$ in this limit. As it was seen, the solution
for $\alpha=0$ and $\gamma/\alpha=4$ does not have a problem in the
$T\rightarrow\infty$ limit. For the other solutions, one has to go
back to Eq. (\ref{Passim-dif4}) and analyze the pressure. As $m/T$
is proportional to powers of $g_s$, the terms in $m/T$ do not
disagree with the ideal massless gas limit. On the other hand, the
term involving $b_0\alpha$ does. The only way to solve this problem
is to make $b_0\alpha$ term vanish, what implies that the powers of
$T$ must be negative. Hence, the general solution does not make
physical sense for $\gamma/\alpha>4$. As a result, the general
solution obtained in section II has possible physical meaning only
if the condition $\gamma/\alpha\leq4$ is satisfied. We will call
this condition of \textit{weak physical condition}. Notice that
solutions $1$, $2$ and $3$ satisfy the weak physical condition.

\section{Analyzing the Physics Solutions}

We start with solution $1$ at very high temperatures. The pressure
in this scenario will be denoted by $P_1$, and can be written with
the help of Eq. (\ref{Passim-dif4}) and the condition $\gamma=0$ as:

\ba
P_1(T)=P_{0}\left[1-\frac{15}{8\pi^2}\le(\frac{m}{T}\ri)^2+\frac{15}{8\pi^3}\le(\frac{m}{T}\ri)^3+\ldots\ri].
\label{P_1m} \ea

\ni Introducing the corrected asymptotic mass, Eq. (\ref{m/T}), in
Eq. (\ref{P_1m}), one obtains

\be
P_1=P_{0}\left[1-\frac{15}{16\pi^2}g_s^2+\frac{15}{4\pi^3}g_s^3\left(\frac{3\sqrt{2}+1}{4\sqrt{2}}\right)+\ldots\right].
\label{P_1} \ee

On the other hand, the perturbative QCD expression for the
pressure at high temperature\cite{kapusta}, known so far up to
order $g_s^6\log(1/g_s)$\cite{kajantie}, is:

\be
P_{QCD}=P_0\left[1-\frac{15}{16}\frac{g_s^2}{\pi^2}+\frac{15}{4}\frac{g_s^3}{\pi^3}+\ldots\right].
\label{p-qcd} \ee

\ni Comparing Eqs. (\ref{P_1}) and (\ref{p-qcd}), one sees that the
zero and second order terms match, but the $g_s^3$ term does
not\footnote{The number inside the parenthesis which multiplies
$g_s^3$ in the Eq. (\ref{P_1}) is $\sim 0.93$.}. Even so, besides
the weak physical condition being satisfied, and the matching
between asymptotic pressure and perturbative QCD up to $g_s^2$
order, solution 1 has another quality: the connection between
thermodynamics and statistical mechanics can be written in such a
way which preserve the same form of standard statistical mechanics,
Eq. (\ref{finalexp-SUA}) with $\gamma=0$.

One has to remember that the extra term $B$ is usually interpreted
as a modification in the zero point energy, in other words, the
vacuum is modified. If the vacuum changes, and now it is $T$
dependent, it is expected an entropy associated with this vacuum,
and consequently a $T$ dependent vacuum entropy. But solution 1 does
not have an extra term associated to a vacuum entropy, which can be
seen as a limitation of this particular solution.

The pressure for solution $2$ at extremely high temperatures,
denoted by $P_2$, is:

\ba
P_2=P_{0}\left[1-\frac{15}{12\pi^2}\le(\frac{m}{T}\ri)^2+O(m^4/T^4)+\ldots\right].\nonumber
\ea

\ni Notice that there is not a correction in $m^3/T^3$. This happens
because the extra term $B$ cancels out the term coming from the SM
pressure. If one uses the HTL asymptotic mass in the pressure $P_2$,
then

\ba
P_2=P_{0}\left[1-\frac{15}{24}\frac{g_s^2}{\pi^2}+\frac{15}{24}\frac{g_s^3}{\pi^3}+\ldots\right].\label{P_2}
\ea

\ni The pressure $P_2$ does not match with pQCD at any order in
$g_s$. The term $g_s^2$ is about $1.5$ smaller than pQCD, while the
$g_s^3$ is 6 times smaller. However, there is one way for solution 2
to match the pQCD result. If one uses $m^2=3g_s^2T^2/4$ for the
mass, instead of the HTL asymptotic mass, the perturbative result is
recovered. In Ref. \cite{Bannur}, the gluon mass was taken as the
plasma frequency $m^2=\omega^2=\frac{2N_cg_s^2T^2}{18}$. The effect
of using the plasma frequency leaves the pressure more convergent
than $P_2$ and $P_{QCD}$.

As it was seen in section $3.1$, solution $3$ is the only solution
which has algebraic expressions for the thermodynamics functions.
This solution, asymptotically, has a pressure $P_3$ given by:

\be
P_{3}=P_0\le[1-\frac{15}{8}\frac{g_s^2}{\pi^2}+\frac{15}{8}\frac{g_s^3}{\pi^3}(3+\sqrt{2})
+\ldots\right]. \nonumber \ee

Comparing the asymptotic behavior of the $P_3$, calculated with the
HTL asymptotic mass, with the perturbative QCD pressure, one sees
that there is no match at any order, as expected. However, if
instead of the HTL mass one uses $m=\frac{g_sT}{2}$ for the mass,
the QCD pressure up to $g_s^2$ is recovered. Motivated by the HTL
gap equation, one can modify the mass relation in general and try a
matching also at higher orders. One way to do this is dividing Eq.
(\ref{m-NLO}) by two, and redefining $\frac{m}{\sqrt{2}}\rightarrow
m$. At lowest order the mass will be $g_sT/2$. Extending the mass
equation to second order, the $g_s^3$ term for the pressure will
match the perturbative QCD result well. Hence, as a bonus, solution
3 with a modified gap equation for the thermal mass, results in a
match between  $P_3$ and $P_{QCD}$ up to the $g_s^3$ order.

As the internal energy and entropy are changed by a $T$ dependent
function in solution 3, the extra $B$ term can be regarded as a
zero-point energy, with an entropy associated with it. It is
important to emphasize that as one has already obtained an algebraic
expression for the pressure at the whole temperature range, it is
possible to fit $m(T)$ by QCD lattice data, and search for
deviations of the HTL asymptotic mass near of $T_c$.

The solution $\gamma/\alpha=4$ has also to be analyzed. This
solution is the only one which has a term of order $g_s$. Although
it satisfies the weak physical condition, the contribution of order
$g_s$ contradicts any pressure computation in the literature, either
HTL, 2-Loop $\Phi$-derivable approximation or perturbative QCD. The
absence of a term in $g_s$ can be justified by thermal field theory,
since the first-order correction to the partition function is
proportional to $g_s^2$ \cite{lebellac}. Therefore, we will let this
solution out of the possible physical solutions.


\section{Discussion and Summary}

In this work we have studied the thermodynamics and the statistical
mechanics of a system with a temperature dependent Hamiltonian in
the canonical ensemble. Earlier works have shown that a system with
a $T$ dependent Hamiltonian is thermodynamically inconsistent if the
connection between thermodynamics and statistical mechanics is the
same as in the $T$ independent Hamiltonian case. In order to have a
general solution to this problem, we developed a formalism which
gives a general connection between thermodynamics and statistical
mechanics, where all the thermodynamics relations are satisfied. We
have seen that it is necessary to add an extra term to the
thermodynamics relations to guarantee the thermodynamics
consistency, and depending on the choice that is made for the
parameters $\alpha$, $\gamma$ and $\eta$, the extra term $B$ can be
simpler.

The general solution developed here is a generalization of the
solution proposed by Gorenstein and Yang \cite{go.1}. In their work,
they have introduced one particular manner to maintain the
thermodynamics consistency: a correction coming from the zero point
energy of the Hamiltonian modifies the pressure and the internal
energy of the system, while leaving the entropy unchanged. In the
present work, it is shown that letting the Hamiltonian function to
be dependent on the temperature of the system, represented by a
function $f(T)$, implies that there are, in principle, a large
number of ways to render the thermodynamics of the system
consistent. We emphasize that this procedure is applicable in any
system with a $T$ dependent Hamiltonian, and not only in the case of
quasi-particle models.

For an ideal quasi-gluon gas, we computed the pressure, the energy
and the entropy, within the formalism developed in section $2$. This
calculation, named solution 3, is easier to work with as no integral
in $T$ is necessary to determine the extra term $B$. As the
formalism does not provide an expression for the thermal mass
$m(T)$, we used the mass calculated in the HTL approach. We achieved
an optimal fit for the thermodynamics functions with two free
parameters in the expression for the strong coupling at finite $T$.

We also studied the general asymptotic behavior of the solutions
contained in $\alpha\neq 0$ cases. The analysis of these solutions
resulted in a reduced number of possible physical solutions. The
main properties of solutions 1, 2, and 3 were discussed, and it was
showed that solution 1 is the only one of the three solutions which
recovers pQCD to order $g_s^2$ when the HTL asymptotic mass is used.
In order for solutions 2 and 3 recover pQCD, it is necessary to
change the gap equation for the mass. Finally, we saw that the only
solutions that have physics meaning are those where the constants
$\alpha$, $\gamma$ and $\eta$ satisfy the weak physical condition,
$\alpha=0$ or $\gamma/\alpha<4$, together with the constraint
$\alpha=\gamma+\eta$. An extension of the present calculation to the
case of a finite chemical potential will be presented in the near
future.
\\
\\
This work was supported by FAPESP (04/15276-2) and CNPq
(307284/2006-9).


\begin{appendix}

\section{Appendix}

We need results for integrals like \be
I_n(r)=\frac{1}{\Gamma(n)}\int_{0}^{\infty}dx\frac{x^{n-1}}{(x^2+r^2)^{\frac{1}{2}}}\frac{1}{e^{(x^2+r^2)^{\frac{1}{2}}}-1}.
\label{I}\ee

\ni The relevant integrals for our calculation, following
\cite{kapusta,jack} and generalizing their results, are the
following:

\be
I_3(r)=\frac{\pi^2}{2^23}-\frac{\pi}{2^2}r-\frac{r^2}{2^3}\left(\log\frac{r}{2^2\pi}+\gamma_E-\frac{1}{2}\right)-\frac{1}{2^2}\sum^{\infty}_{m=1}
a_m(m+2)r^{2(m+1)}. \label{I3} \ee

\be I_5(r)=\frac{\pi^4}{2^33^25}-r^2
\frac{\pi^2}{2^53}+\frac{\pi}{2^43}r^3+\frac{r^4}{2^7}\left(\log\frac{r}{2^2\pi}+\gamma_E-\frac{3}{2^2}\right)
+\frac{1}{2^5}\sum^{\infty}_{m=1}a_mr^{2(m+2)}. \label{I5} \ee

\ni where
$a_m=\frac{(-1)^m(2m-1)!!\zeta(2m+1)}{2^{3m+1}\pi^{2m}(m+2)!!}$.
If one tests the series for convergence, for instance the ratio
test, one finds that the series are convergent for $r<2\pi$.

\end{appendix}

\end{document}